\documentclass{kapproc}

\usepackage[dvips]{graphicx}
\normallatexbib

\newcommand{\Msun}{{\rm M}_{\odot}}

\begin{document}

\articletitle[Formation of globular clusters]
             {First Starbursts at high redshift: \\
              Formation of globular clusters}

\author{Oleg Y. Gnedin}

\affil{Space Telescope Science Institute}
\email{ognedin@stsci.edu}

\begin{abstract}  
  Numerical simulations of a Milky Way-size galaxy demonstrate that
  globular clusters with the properties similar to observed can form
  naturally at $z>3$ in the concordance $\Lambda$CDM cosmology.  The
  clusters in our model form in the strongly baryon-dominated cores of
  supergiant molecular clouds.  The first clusters form at $z \approx 12$,
  while the peak formation appears to be at $z \sim 3-5$.  The
  zero-age mass function of globular clusters can be approximated by a
  power-law $dN/dM \propto M^{-2}$ in agreement with observations of
  young massive star clusters.
\end{abstract}

For the first time it has been possible to include the formation of
globular clusters self-consistently into the hierarchical galaxy
formation model \cite{kravtsov_gnedin03}.  The simulations used in our
study were performed using the Eulerian gasdynamics$+N$-body ART code
which uses an adaptive mesh refinement approach to achieve high
dynamic range \cite{kravtsov_etal97}.  Several physical processes
critical to various aspects of galaxy formation are included: star
formation; metal enrichment and thermal feedback due to supernovae
type II and type Ia; self-consistent advection of metals; metallicity-
and density-dependent cooling and UV heating due to the cosmological
ionizing background.  The simulation follows the early ($z > 3$)
stages of the evolution of a Milky Way-type galaxy, $M \approx
10^{12}h^{-1}\ \rm M_{\odot}$ at $z=0$.

Although the resolution achieved in the disk region is very high by
cosmological standards, $\Delta x \approx 50$ pc at $z=4$, it is still
insufficient to resolve the formation of stellar clusters. The
resolution, however, is sufficient to identify the potential sites for
GC formation.  The natural candidates are cores of giant molecular
clouds in high-redshift galaxies \cite{harris_pudritz94}.
Accordingly, we complement the simulations with a physical description
of the gas distribution on a subgrid level.  It assumes an isothermal
structure of the unresolved core and a universal density threshold of
cluster star formation, $\rho_{\rm sf} = 10^4 \, \Msun$ pc$^{-3}$.
The gas at densities above $\rho_{\rm sf}$ is converted into a star
cluster with the efficiency $\epsilon = M_*/M_{\rm core} = 60\%$.  The
size of the cluster after the gradual loss of the remaining gas is
$R_* = R_{\rm core}/\epsilon$.

\begin{figure}[t]
\begin{center}
\includegraphics[width=.75\textwidth]{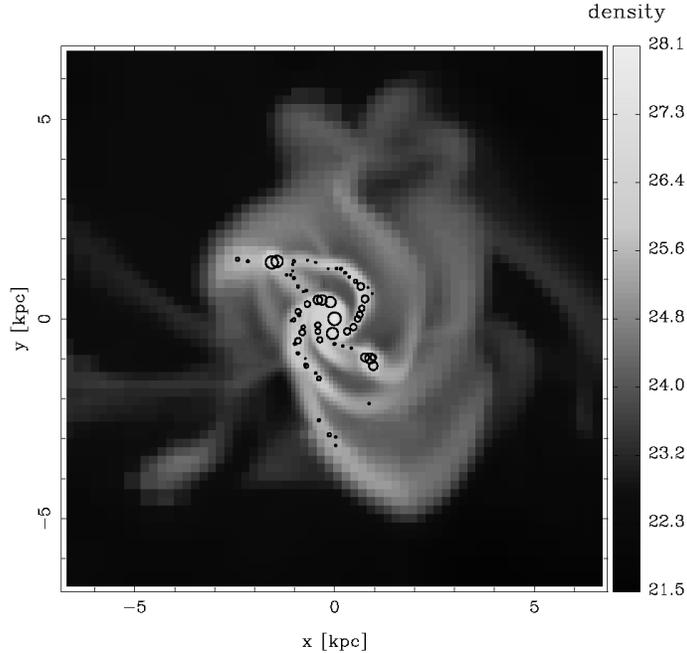}
\end{center}
\caption{Projected gas density in the most massive disk at $z=4$.  A
  nearly face-on disk has prominent spiral arms and is in the process of
  very active accretion and merging.  The globular clusters identified
  at this epoch are shown by circles.  \label{fig:dtv}}
\end{figure}

{\em The first globular clusters formed around} $z \approx 12$ and
cluster formation continued at least until $z \approx 3$. The
preferred epoch of globular cluster formation ($z \sim 3-5$)
corresponds to the cosmic time of only $1-2$~Gyr, when the gas supply
is abundant in the disks of the progenitor halos and the merger rate
of the progenitors is high.  All old globular clusters thus appear to
have similar ages.

Most globular clusters in our model form in halos of mass $> 10^9 \
\Msun$.  {\em Within the progenitor systems, globular clusters form in
the highest-density regions of the disk.}  In the most massive disk in
the simulation (Figure~\ref{fig:dtv}) the newly formed clusters trace
the spiral arms and the nucleus, similarly to the young star clusters
observed in merging and interacting galaxies \cite{whitmore_etal99}.

Note that although the high-redshift globular clusters form in gaseous
disks, the subsequent accretion of their parent galaxies would lead to
tidal stripping and disruption.  The clusters are likely to share the
fate of the stripped stars that build up the galactic stellar halo and
should have roughly {\it spherical spatial distribution at} $z=0$.

\begin{figure}[t]
\begin{center}
\includegraphics[width=.6\textwidth]{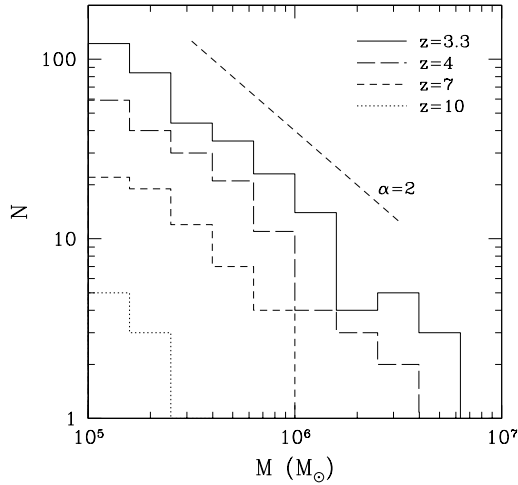}
\end{center}
\caption{The build-up of the initial mass function of globular
  clusters at four epochs.  The straight dashed line shows a power-law,
  $N\propto M^{-2}$.  The slope $\alpha$ becomes shallower with
  decreasing redshift and saturates at $\alpha = 2.05 \pm 0.07$ for
  $z < 4$.  \label{fig:gc_m}}
\end{figure}

The total mass of clusters within a parent galactic halo correlates
with the total galaxy mass $M_{\rm h}$: $M_{\rm GC} = 3.2 \times 10^6
\, \Msun \left({M_{\rm h} / 10^{11} \, \Msun}\right)^{1.13 \pm 0.08}$.
The global efficiency of cluster formation, $M_{\rm GC}/M_{\rm h}$,
therefore depends only weakly on the galaxy mass.  The mass of the
globular cluster population in a given region also strongly correlates
with the local average star formation rate density: $M_{\rm GC}\propto
\Sigma_{\rm SFR}^{0.75\pm 0.06}$ at $z=3.3$.  A similar correlation,
albeit with a significant scatter, was reported for the observed
present-day galaxies \cite{larsen02}.  In our model the correlation
arises because both the star formation rate and mass of the globular
cluster population are controlled by the same parameter, the amount of
gas in the densest regions of the ISM.

The metallicities, which the clusters acquire from the gas in which
they form, are remarkably similar to the metal-poor part of the
Galactic cluster distribution.  Note that at all epochs the dynamical
time of the molecular cores is very short ($\sim 10^6$~yrs), which
means that the galactic gas is pre-enriched even before the first
clusters form.  Thus {\it the oldest globular clusters do not contain
the oldest stars in the Galaxy}.

The mass function of the model clusters at all output epochs (Figure
\ref{fig:gc_m}) is well fit by a power-law $dN/dM \propto M^{-2}$ for
$M > 10^5 \, \Msun$, in agreement with the mass function of {\em
young} star clusters in normal spiral and interacting galaxies
\cite{zhang_fall99}.  The power-law slope at high masses also
resembles that of the high-redshift mass function of dark matter halos
in the hierarchical CDM cosmology \cite{gnedin03}.  We find that
although the most massive cluster in each galaxy correlates with the
parent galaxy mass, $M_{\rm max} \propto M_{\rm h}^{1.3 \pm 0.1}$, a
significant scatter around this relation implies that for a given halo
the masses of individual clusters can vary by a factor of three.
Therefore, the {\em overall shape of the mass function of globular
clusters in our model does not follow directly from the mass function
of their parent halos, but depends also on the mass function of
molecular cloud cores within individual galaxies}.

The slope of the mass function steepens with increasing mass.  We find
a relatively shallow, $\alpha \approx 1.7$, mass function for the
small-mass clusters forming in lower-density cores and the steep,
$\alpha \approx 2$, mass function for massive globular clusters
forming in the densest regions of the disk.  The steepening of the
luminosity function with increasing luminosity is also observed for
young star clusters in starbursting galaxies
\cite{whitmore_etal99,larsen02}.  {\em Thus over the range of masses
typically probed in observations, $M > 10^5 \ \Msun$, the mass
function can be approximated by a power-law because the expected
curvature over this range is rather small.}

\chapbblname{gnedin}
\chapbibliography{gc}

\end{document}